\documentclass[twocolumn,showpacs,preprintnumbers,amsmath,amssymb, showkeys]{revtex4}

\usepackage{graphicx}
\usepackage{dcolumn}
\usepackage{bm}
\usepackage{epsf}

\begin{document}


\title{The effect of Coulomb correlations on non-equilibrium charge redistribution tuned by the tunneling current}

\author{P.\,I.\,Arseyev}
 \altaffiliation{ars@lpi.ru}
\author{N.\,S.\,Maslova}%
 \email{spm@spmlab.phys.msu.ru}
\author{V.\,N.\,Mantsevich}
 \altaffiliation{vmantsev@spmlab.phys.msu.ru}
\affiliation{%
 P.N. Lebedev Physical institute of RAS, 119991, Moscow, Russia\\~\\
 Moscow State University, Department of  Physics,
119991 Moscow, Russia
}%

\date{\today }
12 pages, 8 figures
\begin{abstract}
It was shown that tunneling current flowing through a system with
Coulomb correlations leads to charge redistribution between the
different localized states. Simple model consisting of two electron
levels have been analyzed by means of Heisenberg motion equations
taking into account all order correlations of electron filling
numbers in localized states exactly. We consider various relations
between Coulomb interaction and localized electron energies.  Sudden
jumps of electron density at each level in a certain range of
applied bias have been found. We found that for some parameter range
inverse occupation in the two-level system appeared due to Coulomb
correlations. It was shown also that Coulomb correlations lead to
appearance of negative tunneling conductivity at certain relation
between the values of tunneling rates from the two electronic
levels.
\end{abstract}

\maketitle

\section{Introduction}

 Non-equilibrium Coulomb correlations can drastically influence on
the local charge distribution in the vicinity of impurity complexes
in nanometer tunneling junctions. Coulomb interaction results in
significant changes of each localized state electron filling numbers
and current-voltage characteristics of impurity complexes. Adjusting
parameters of a tunneling contact one can obtain negative tunneling
conductivity caused by Coulomb correlations in a certain range of
applied bias. There are several experimental situations in which
Coulomb interaction values are of the order of electron levels
spacing or even strongly exceed this value. It usually takes place
if the distance between several impurity atoms or surface defects is
comparable with the lattice scale, so coupling between their
electronic states can strongly exceeds the interaction of these
localized states with continuous spectrum. Another possible
realization is a quantum dot or two small interacting quantum dots
on the sample surface weakly connected with the bulk states. Such
systems can be described by the model including several electron
levels with Coulomb interaction between localized electrons.
Electronic structure of such complexes can be tuned both by external
electric field which changes the values of single particle levels
and by electron correlations of localized electronic states. One can
expect that Coulomb correlations in non equilibrium situation result
in spatial redistribution of localized charges and possibility of
local charge density manipulation governed by Coulomb correlations.
In some sense these effects are similar to the $"$co-tunneling$"$
observed in \cite{Feigel'man}, \cite{Beloborodov}. Moreover Coulomb
interaction of localized electrons can be responsible for inverse
occupation of localized electron states and negative local tunneling
conductivity in a certain range of applied bias. These effects can
be clearly seen if single electron levels have different spatial
symmetry.

Great attention was paid to electron transport through a single
impurity or a dot in the Coulomb blockade and the Kondo \cite{Kondo}
regimes. These effects have been studied experimentally and are up
till now under theoretical investigation
\cite{Goldin}-\cite{Kikoin}.  But if tunneling coupling is not
negligible the impurity charge is not a discrete value and one has
to deal with impurity electron filling numbers (which now are
continuous variables) determined from kinetic equations.

Non-equilibrium effects and tunneling current spectra in the system
of two weakly coupled impurities (when coupling between impurities
is smaller than tunneling rates between energy levels and tunneling
contact leads) in the presence of Coulomb interaction were described
by self-consistent approach based on Keldysh diagram technique in
\cite{Keldysh},\cite{Arseyev}. In the present work we consider the
opposite case when Coulomb coupling between localized electron
states strongly exceeds tunneling transfer rates.

We suggest theoretical approach based on the Heisenberg equations
for localized states electron filling numbers taking into account
all order correlators of local electron density \cite{Maslova}.
Tunneling current in a two-level system of spinless fermions with
infinite value of Coulomb interaction has been investigated in
\cite{Kuznetsov}. But obtained results do not take into account any
nontrivial pair correlations in the system for finite Coulomb
correlations. If one is interested in kinetic properties for the
applied bias range larger than the characteristic energy of
correlations between localized and band electrons in the leads then
Kondo effect is unimportant. In this case for the finite number of
localized electron levels one can obtain closed system of equations
for electron filling numbers and all their correlators. It allows to
analyze the role of Coulomb correlations in charge redistribution
and in formation main features of I-V characteristics.

\section{The suggested model}

\begin{figure} [h]
\centering
\includegraphics[width=50mm]{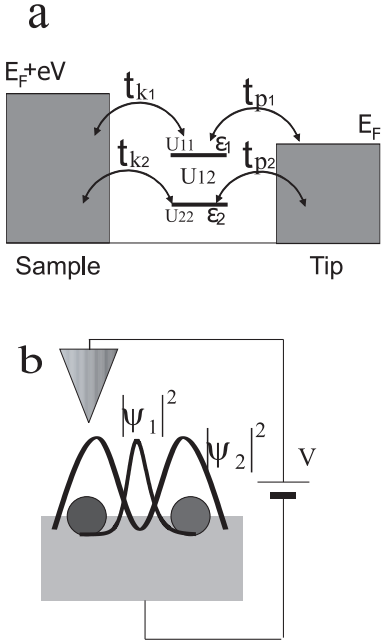}
\caption{ a). Energy diagram of two-level system and b). Schematic
spatial diagram of experimental realization. Coulomb energy $U_{ij}$
correspond to the interaction between electrons on different energy
levels.} \label{Fig.1}
\end{figure}

We shall analyze tunneling through the two-level system with Coulomb
interaction of localized electrons Fig.\ref{Fig.1}. The model system
can be described by the Hamiltonian $\hat{H}$.

\begin{eqnarray}
\hat{H}=\sum_{i\sigma}\varepsilon_{i}n_{i\sigma}+\sum_{k\sigma}\varepsilon_{k}c_{k\sigma}^{+}c_{k\sigma}+\nonumber\\
+\sum_{p\sigma}\varepsilon_{p}c_{p\sigma}^{+}c_{p\sigma}
+\sum_{ij\sigma\sigma^{'}}U_{ij}^{\sigma\sigma^{'}}n_{i\sigma}n_{j\sigma^{'}}+\nonumber\\
+\sum_{ki\sigma}t_{ki}(c_{k\sigma}^{+}c_{i\sigma}+h.c.)+\sum_{pi\sigma}t_{pi}(c_{p\sigma}^{+}c_{i\sigma}+h.c.)
\end{eqnarray}

Indices $k$ and $p$ label continuous spectrum states in the left
(sample) and right (tip) leads of tunneling contact respectively.
$t_{k(p)}$- tunneling transfer amplitudes between continuous
spectrum states and localized states with energies $\varepsilon_i$.
 Operators
$c_{k(p)}^{+}/c_{k(p)}$ correspond to electrons
creation/annihilation in the continuous spectrum states $k(p)$.
$n_{i\sigma}=c_{i\sigma}^{+}c_{i\sigma}$-two-level system electron
filling numbers, where operator $c_{i\sigma}$ destroys electron with
spin $\sigma$ on the energy level $\varepsilon_i$.

Tunneling current through the two-level system is written in terms
of electron creation/annihilation operators as:

\begin{eqnarray}
I=I_{k\sigma}=\sum_{i\sigma}I_{ki\sigma}=\sum_{k\sigma}\dot{n}_{k\sigma}=\nonumber\\
=\sum_{ki\sigma}t_{ki}(<c_{k\sigma}^{+}c_{i\sigma}>-<c_{i\sigma}^{+}c_{k\sigma}>)
\end{eqnarray}

Let us consider $\hbar=1$ elsewhere, so motion equation for the
electron operators product $c_{k\sigma}^{+}c_{i\sigma}$ can be
written as:

\begin{eqnarray}
i\frac{\partial c_{k\sigma}^{+}c_{i\sigma}}{\partial
t}=(\varepsilon_i-\varepsilon_k)\cdot
c_{k\sigma}^{+}c_{i\sigma}+U_{ii}n_{i-\sigma}\cdot
c_{k\sigma}^{+}c_{i\sigma}+\nonumber\\
+U_{ij}(n_{j\sigma}+n_{j-\sigma})\cdot c_{k\sigma}^{+}c_{i\sigma}
-t_{ki}\cdot (n_{i\sigma}-\widehat{f}_{k})+\nonumber\\
+\sum_{k^{'}\neq
k}t_{k^{'}i}c_{k\sigma}^{+}c_{k^{'}\sigma}+\sum_{i\neq
j}t_{kj}c_{j\sigma}^{+}c_{i\sigma}=0 \label{1}
\end{eqnarray}

where

\begin{eqnarray}
\widehat{f}_{k}=c_{k\sigma}^{+}c_{k\sigma}
\end{eqnarray}

In order to get equation for the tunneling current one has to
multiply equation (\ref{1}) by combinations of electron filling
numbers operators $n_{i(j)\pm\sigma}$ in the following way:

\begin{eqnarray}
(1-n_{1-\sigma})(1-n_{2-\sigma})(1-n_{2\sigma})c_{k\sigma}^{+}c_{1\sigma}=\nonumber\\
=\{(t_{k1}\cdot(n_{1\sigma}-\widehat{f}_{k})+\sum_{k^{'}\neq k}t_{k^{'}1}c_{k\sigma}^{+}c_{k^{'}\sigma}+t_{k2}c_{2\sigma}^{+}c_{1\sigma})\cdot\nonumber\\
\cdot(1-n_{1-\sigma})(1-n_{2-\sigma})(1-n_{2\sigma})\}\cdot\{\varepsilon_{1}-\varepsilon_{k}\}^{-1}\nonumber\\
\label{eq_1}
\end{eqnarray}
\begin{eqnarray}
n_{1-\sigma}(1-n_{2-\sigma})(1-n_{2\sigma})c_{k\sigma}^{+}c_{1\sigma}=\nonumber\\
=\{(t_{k1}\cdot(n_{1\sigma}-\widehat{f}_{k})+\sum_{k^{'}\neq k}t_{k^{'}1}c_{k\sigma}^{+}c_{k^{'}\sigma}+t_{k2}c_{2\sigma}^{+}c_{1\sigma})\cdot\nonumber\\
\cdot n_{1-\sigma}(1-n_{2-\sigma})(1-n_{2\sigma})\}\cdot\{\varepsilon_{1}-\varepsilon_{k}+U_{11}\}^{-1}\nonumber\\
\end{eqnarray}
\begin{eqnarray}
\sum_{\sigma^{'}}n_{2\sigma^{'}}(1-n_{1-\sigma})(1-n_{2-\sigma^{'}})c_{k\sigma}^{+}c_{1\sigma}=\nonumber\\
=\sum_{\sigma^{'}}\{(t_{k1}\cdot(n_{1\sigma}-\widehat{f}_{k})+\sum_{k^{'}\neq k}t_{k^{'}1}c_{k\sigma}^{+}c_{k^{'}\sigma}+t_{k2}c_{2\sigma}^{+}c_{1\sigma})\cdot\nonumber\\
\cdot n_{2\sigma^{'}}(1-n_{1-\sigma})(1-n_{2-\sigma^{'}})\}\cdot\{\varepsilon_{1}-\varepsilon_{k}+U_{12}\}^{-1}\nonumber\\
\end{eqnarray}
\begin{eqnarray}
\sum_{\sigma^{'}}n_{1-\sigma}n_{2\sigma^{'}}(1-n_{2-\sigma^{'}})c_{k\sigma}^{+}c_{1\sigma}=\nonumber\\
=\sum_{\sigma^{'}}\{(t_{k1}\cdot(n_{1\sigma}-\widehat{f}_{k})+\sum_{k^{'}\neq
k}t_{k^{'}1}c_{k\sigma}^{+}c_{k^{'}\sigma}+t_{k2}c_{2\sigma}^{+}c_{1\sigma})\cdot\nonumber\\
\cdot n_{1-\sigma}n_{2\sigma^{'}}(1-n_{2-\sigma^{'}})\}\cdot\{\varepsilon_{1}-\varepsilon_{k}+U_{11}+U_{12}\}^{-1}\nonumber\\
\end{eqnarray}
\begin{eqnarray}
n_{2-\sigma}n_{2\sigma}(1-n_{1-\sigma})c_{k\sigma}^{+}c_{1\sigma}=\nonumber\\
=\{(t_{k1}\cdot(n_{1\sigma}-\widehat{f}_{k})+\sum_{k^{'}\neq k}t_{k^{'}1}c_{k\sigma}^{+}c_{k^{'}\sigma}+t_{k2}c_{2\sigma}^{+}c_{1\sigma})\cdot\nonumber\\
\cdot n_{2-\sigma}n_{2\sigma}(1-n_{1-\sigma})\}\cdot\{\varepsilon_{1}-\varepsilon_{k}+2U_{12}\}^{-1}\nonumber\\
\end{eqnarray}
\begin{eqnarray}
n_{1-\sigma}n_{2-\sigma}n_{2\sigma}c_{k\sigma}^{+}c_{1\sigma}=\nonumber\\
=\{(t_{k1}\cdot(n_{1\sigma}-\widehat{f}_{k})+\sum_{k^{'}\neq k}t_{k^{'}1}c_{k\sigma}^{+}c_{k^{'}\sigma}+t_{k2}c_{2\sigma}^{+}c_{1\sigma})\cdot\nonumber\\
\cdot n_{1-\sigma}n_{2-\sigma}n_{2\sigma}\}\cdot\{\varepsilon_{1}-\varepsilon_{k}+U_{11}+2U_{12}\}^{-1}\nonumber\\
\label{eq_2}
\end{eqnarray}

The relation $n_{i\sigma}^{2}=n_{i\sigma}$ was used in these
equations.

Neglecting changes of electron spectrum and local density of states
in the tunneling contact leads caused by the tunneling current we
shall uncouple conduction and localized electron filling numbers.
This means also that we neglect any correlation effects between
localized and band electrons - like the Kondo effect.

It is easy to check:

\begin{eqnarray}
(1-n_{1-\sigma})(1-n_{2-\sigma})(1-n_{2\sigma})+\nonumber\\+n_{1-\sigma}(1-n_{2-\sigma})(1-n_{2\sigma})+\nonumber\\
+\sum_{\sigma^{'}}n_{2\sigma^{'}}(1-n_{1-\sigma})(1-n_{2-\sigma^{'}})+\nonumber\\+\sum_{\sigma^{'}}n_{1-\sigma}n_{2\sigma^{'}}(1-n_{2-\sigma^{'}})+\nonumber\\
+n_{2-\sigma}n_{2\sigma}(1-n_{1-\sigma})+n_{1-\sigma}n_{2-\sigma}n_{2\sigma}
=1
\end{eqnarray}

Thus summing up the right- and left-hand parts of equations
(\ref{eq_1}-\ref{eq_2}) we get an equation for
$<c_{k\sigma}^{+}c_{i\sigma}>$, which gives us after summation over
$k$ an equation for the tunneling current through the two-level
system. Total current is a sum of two contributions:

\begin{eqnarray}
I_{k\sigma}=I_{k1\sigma}+I_{k2\sigma}
\end{eqnarray}

Where expression for the tunneling current $I_{k2\sigma}$ can be
obtained by changing indexes $1\leftrightarrow2$ in the equation for
the tunneling current $I_{k1\sigma}$, which has the following form:

\begin{eqnarray}
I_{k1\sigma}&=&\Gamma_{k1}\{\langle n_{1\sigma}\rangle-\langle(1-n_{1-\sigma})(1-n_{2-\sigma})(1-n_{2\sigma})\rangle f_{k}(\varepsilon_1)-\nonumber\\
&-&\langle n_{1-\sigma}(1-n_{2-\sigma})(1-n_{2\sigma})\rangle\cdot f_{k}(\varepsilon_1+U_{11})-\nonumber\\
&-&\langle n_{2\sigma}(1-n_{2-\sigma})(1-n_{1-\sigma})\rangle\cdot f_{k}(\varepsilon_1+U_{12})-\nonumber\\
&-&\langle n_{2-\sigma}(1-n_{2\sigma})(1-n_{1-\sigma})\rangle\cdot f_{k}(\varepsilon_1+U_{12})-\nonumber\\
&-&\langle n_{1-\sigma}n_{2\sigma}(1-n_{2-\sigma})\rangle\cdot f_{k}(\varepsilon_1+U_{11}+U_{12})-\nonumber\\
&-&\langle n_{1-\sigma}n_{2-\sigma}(1-n_{2\sigma})\rangle\cdot f_{k}(\varepsilon_1+U_{11}+U_{12})-\nonumber\\
&-&\langle n_{2\sigma}n_{2-\sigma}(1-n_{1-\sigma})\rangle\cdot f_{k}(\varepsilon_1+2U_{12})-\nonumber\\
&-&\langle n_{1-\sigma}n_{2-\sigma}n_{2\sigma}\rangle\cdot
f_{k}(\varepsilon_1+U_{11}+2U_{12})\}+\nonumber\\
&+&t_{k1}t_{k2}\nu_{0k}c_{2\sigma}^{+}c_{1\sigma}+\nonumber\\
&+&\sum_{k^{'}\neq
k}\langle t_{k1}t_{k^{'}1}c_{k\sigma}^{+}c_{k^{'}\sigma}\rangle\cdot\nonumber\\
&\cdot&\{\langle\frac{(1-n_{1-\sigma})(1-n_{2-\sigma})(1-n_{2\sigma})}{\varepsilon_1-\varepsilon_k}\rangle+\nonumber\\
&+&\langle\frac{n_{1-\sigma}(1-n_{2-\sigma})(1-n_{2\sigma})}{\varepsilon_1+U_{11}-\varepsilon_k}\rangle+\nonumber\\
&+&\langle\frac{\sum_{\sigma^{'}}n_{2\sigma^{'}}(1-n_{1-\sigma})(1-n_{2-\sigma^{'}})}{\varepsilon_1+U_{12}-\varepsilon_k}\rangle+\nonumber\\
&+&\langle\frac{\sum_{\sigma^{'}}n_{1-\sigma}n_{2\sigma^{'}}(1-n_{2-\sigma^{'}})}{\varepsilon_1+U_{11}+U_{12}-\varepsilon_k}\rangle+\nonumber\\
&+&\langle\frac{n_{2-\sigma}n_{2\sigma}(1-n_{1-\sigma})}{\varepsilon_1+2U_{12}-\varepsilon_k}\rangle+\langle\frac{n_{1-\sigma}n_{2-\sigma}n_{2\sigma}}{\varepsilon_1+U_{11}+2U_{12}-\varepsilon_k}\rangle\}\
\label{current}
\end{eqnarray}

In what follows we shall neglect terms
$t_{k1}t_{k2}\nu_{0k}c_{2\sigma}^{+}c_{1\sigma}$ and terms
proportional
$\frac{t_{k1}t_{k^{'}1}c_{k\sigma}^{+}c_{k^{'}\sigma}}{\varepsilon_1-\varepsilon_k}$
in the expression (\ref{current}) as they correspond to the next
order perturbation theory in the parameter
$\frac{\Gamma_{i}}{\varepsilon_{i}}$. Relaxation rates
$\Gamma_{k(p)i}=\pi\cdot t_{k(p)i}^{2}\cdot\nu_{0}$ are determined
by electron tunneling transitions from the two-level system to the
leads $k$ (sample) and $p$ (tip) continuum states.
$\nu_{0k(p)}$-continuous spectrum density of states. The main
equation for the current (\ref{current}) includes mean electron
filling numbers $n_{i\sigma}$, pair and triple correlators for the
localized states, which have to be determined now. Equations for the
total electron filling numbers $n_{1\sigma}$ è $n_{2\sigma}$ on the
levels $1$ and $2$ can be found from the conditions:

\begin{eqnarray}
\frac{\partial n_{1\sigma}}{\partial t}=I_{k1\sigma}+I_{p1\sigma}=0\nonumber\\
\frac{\partial n_{2\sigma}}{\partial t}=I_{k2\sigma}+I_{p2\sigma}=0\
\label{current_1}
\end{eqnarray}

where tunneling current $I_{p\sigma}$ can be easily obtained from
$I_{k\sigma}$ by changing indexes $k\leftrightarrow p$.

Pair filling numbers correlators can be found in the following way:

\begin{eqnarray}
\langle\frac{\partial n_{i\sigma}n_{j\sigma^{'}}}{\partial
t}\rangle=\langle\frac{\partial n_{i\sigma}}{\partial
t}n_{j\sigma^{'}}\rangle+\langle\frac{\partial
n_{j\sigma^{'}}}{\partial t}n_{i\sigma}\rangle
\end{eqnarray}

Full expressions which determine the system of equations for pair
filling numbers correlators through the higher order correlators in
the stationary case have the form:

\begin{eqnarray}
&\langle&\frac{\partial n_{i\sigma}n_{j\sigma^{'}}}{\partial
t}\rangle=(\Gamma_{ki}+\Gamma_{pi}+\Gamma_{kj}+\Gamma_{pj})\cdot\nonumber\\&\cdot&\langle n_{i\sigma}n_{j\sigma^{'}}\rangle-\nonumber\\
&-&(\Gamma_{ki}f_{k}(\varepsilon_{i}+U_{ij})+\Gamma_{pi}f_{p}(\varepsilon_{i}+U_{ij}))\cdot\nonumber\\&\cdot&\langle n_{j\sigma^{'}}(1-n_{j-\sigma^{'}})(1-n_{i-\sigma})\rangle-\nonumber\\
&-&(\Gamma_{kj}f_{k}(\varepsilon_{j}+U_{ij})+\Gamma_{pj}f_{p}(\varepsilon_{j}+U_{ij}))\cdot\nonumber\\&\cdot&\langle n_{i\sigma}(1-n_{i-\sigma})(1-n_{j-\sigma^{'}})\rangle-\nonumber\\
&-&(\Gamma_{ki}f_{k}(\varepsilon_{i}+U_{ii}+U_{ij})+\Gamma_{pi}f_{p}(\varepsilon_{i}+U_{ii}+U_{ij}))\cdot\nonumber\\&\cdot&\langle n_{i-\sigma}n_{j\sigma^{'}}(1-n_{j-\sigma^{'}})\rangle-\nonumber\\
&-&(\Gamma_{ki}f_{k}(\varepsilon_{i}+2U_{ij})+\Gamma_{pi}f_{p}(\varepsilon_{i}+2U_{ij}))\cdot\nonumber\\&\cdot&\langle n_{j-\sigma^{'}}n_{j\sigma^{'}}(1-n_{i-\sigma})\rangle-\nonumber\\
&-&(\Gamma_{ki}f_{k}(\varepsilon_{i}+U_{ii}+2U_{ij})+\Gamma_{pi}f_{p}(\varepsilon_{i}+U_{ii}+2U_{ij}))\cdot\nonumber\\&\cdot&\langle n_{i-\sigma}n_{j\sigma^{'}}n_{j-\sigma^{'}}\rangle-\nonumber\\
&-&(\Gamma_{kj}f_{k}(\varepsilon_{j}+U_{jj}+U_{ij})+\Gamma_{pj}f_{p}(\varepsilon_{j}+U_{jj}+U_{ij}))\cdot\nonumber\\&\cdot&\langle n_{j-\sigma^{'}}n_{i\sigma}(1-n_{i-\sigma})\rangle-\nonumber\\
&-&(\Gamma_{kj}f_{k}(\varepsilon_{j}+2U_{ij})+\Gamma_{pj}f_{p}(\varepsilon_{j}+2U_{ij}))\cdot\nonumber\\&\cdot&\langle n_{i-\sigma}n_{j\sigma^{'}}(1-n_{j-\sigma^{'}})\rangle-\nonumber\\
&-&(\Gamma_{kj}f_{k}(\varepsilon_{j}+U_{jj}+2U_{ij})+\Gamma_{pj}f_{p}(\varepsilon_{j}+U_{jj}+2U_{ij}))\cdot\nonumber\\&\cdot&\langle
n_{j-\sigma^{'}}n_{i\sigma}n_{i-\sigma}\rangle\}=0\nonumber\\
\label{double}
\end{eqnarray}

High order correlators can be found in the similar way:

\begin{eqnarray}
\langle\frac{\partial
n_{j\sigma}n_{j-\sigma}n_{i-\sigma^{'}}}{\partial
t}\rangle&=&\langle\frac{\partial n_{j\sigma}n_{j-\sigma}}{\partial
t}n_{i-\sigma^{'}}\rangle+\nonumber\\&+&\langle\frac{\partial
n_{i-\sigma^{'}}}{\partial
t}n_{j\sigma}n_{j-\sigma}\rangle\nonumber\\
\end{eqnarray}

So expressions for high order correlations have the form:

\begin{eqnarray}
\langle
n_{j\sigma}n_{j-\sigma}n_{i-\sigma^{'}}\rangle&=&\{\Gamma_{kj}\cdot
f_{k}(\varepsilon_{j}+U_{jj}+2U_{ij})\cdot\nonumber\\
&\cdot&(\langle n_{i-\sigma}n_{j\sigma}\rangle+\langle
n_{i-\sigma}n_{j-\sigma}\rangle)+\nonumber\\
&+&\Gamma_{ki}\cdot f_{k}(\varepsilon_{i}+2U_{ij})\cdot \langle
n_{j\sigma}n_{j-\sigma}\rangle+\nonumber\\&+&\Gamma_{pj}\cdot f_{p}(\varepsilon_{j}+U_{jj}+2U_{ij})\cdot\nonumber\\
&\cdot&(\langle n_{i-\sigma}n_{j\sigma}\rangle+\langle
n_{i-\sigma}n_{j-\sigma}\rangle)+\nonumber\\&+&\Gamma_{pi}\cdot
f_{p}(\varepsilon_{i}+2U_{ij})\cdot \langle
n_{j\sigma}n_{j-\sigma}\rangle\}\cdot\nonumber\\
&\cdot&\{\Gamma_{ki}\cdot\{3+f_{k}(\varepsilon_{i}+2U_{ij})-\nonumber\\&-&f_{k}(\varepsilon_{i}+U_{ii}+2U_{ij})\}+\nonumber\\
&+&\Gamma_{pi}\cdot\{3+f_{p}(\varepsilon_{i}+2U_{ij})-\nonumber\\&-&f_{p}(\varepsilon_{i}+U_{ii}+2U_{ij})\}\}^{-1}\nonumber\\
\label{triple}
\end{eqnarray}

We consider here paramagnetic situation $n_{i\sigma}=n_{i-\sigma}$,
$\langle n_{i\sigma}n_{j\sigma}\rangle=\langle
n_{i\sigma}n_{j-\sigma}\rangle$ and $\langle
n_{i\sigma}n_{i-\sigma}n_{j\sigma}\rangle=\langle
n_{i\sigma}n_{i-\sigma}n_{j-\sigma}\rangle$ (Note that the system of
equations (\ref{current_1})-(\ref{matrix}) allows to analyze
magnetic regime with $n_{i\sigma}\neq n_{i-\sigma}$ as well). So the
system of equations for the pair correlators $K_{11}\equiv \langle
n_{1\sigma}n_{1-\sigma}\rangle$, $K_{22}\equiv \langle
n_{2\sigma}n_{2-\sigma}\rangle$ and $K_{12}\equiv \langle
n_{1\sigma}n_{2\sigma}\rangle$ after substitution of equation
(\ref{triple}) to (\ref{double}) has the form:

\begin{equation}
\left(\begin{array}{ccccc}a_{11} &&  a_{12} && a_{13}\\
a_{21} &&  a_{22} && a_{23}\\
a_{31} &&  a_{32} && a_{33}\end{array}\right)\times
\left(\begin{array}{c}K_{11}\\K_{12}\\K_{22}\end{array}\right)=F
\label{matrix}
\end{equation}

with coefficients $a_{ij}$:

\begin{eqnarray}
a_{11}&=&1\nonumber\\
a_{12}&=&2\cdot
n_{1}^{T}(\varepsilon_1+U_{11})-n_{1}^{T}(\varepsilon_1+U_{11}+U_{12})-\nonumber\\&-&2\cdot\frac{\Gamma_2}{\Gamma_1}\cdot
n_{2}^{T}(\varepsilon_2+U_{22}+U_{12})\cdot \Phi_{1}\nonumber\\
a_{13}&=&-n_{1}^{T}(\varepsilon_1+2\cdot U_{12})\cdot \Phi_{1}\
\end{eqnarray}

\begin{eqnarray}
a_{21}&=&-n_{2}^{T}(\varepsilon_2+2\cdot U_{12})\cdot \Phi_{2}\nonumber\\
a_{22}&=&2\cdot
n_{2}^{T}(\varepsilon_2+U_{22})-n_{2}^{T}(\varepsilon_2+U_{22}+U_{12})-\nonumber\\&-&2\cdot\frac{\Gamma_1}{\Gamma_2}\cdot
n_{1}^{T}(\varepsilon_1+U_{11}+U_{12})\cdot \Phi_{2}\nonumber\\
a_{23}&=&1\
\end{eqnarray}

\begin{eqnarray}
a_{31}&=&\frac{\Gamma_{2}}{\Gamma_{1}+\Gamma_{2}}\cdot (n_{2}^{T}(\varepsilon_2+U_{12})-\nonumber\\&-&n_{2}^{T}(\varepsilon_2+2\cdot U_{12})\cdot(1+2\cdot A_2))\nonumber\\
a_{32}&=&1+\frac{\Gamma_{1}}{\Gamma_{1}+\Gamma_{2}}\cdot
(n_{1}^{T}(\varepsilon_1+U_{12})-\nonumber\\&-&n_{1}^{T}(\varepsilon_1+U_{11}+U_{12})\cdot(1+4\cdot
A_{2}))\nonumber\\
&+&\frac{\Gamma_{2}}{\Gamma_{1}+\Gamma_{2}}\cdot(n_{2}^{T}(\varepsilon_2+U_{12})
-\nonumber\\&-&
n_{2}^{T}(\varepsilon_2+U_{2}+U_{12})\cdot(1+4\cdot A_{1}))\nonumber\\
a_{33}&=&\frac{\Gamma_{1}}{\Gamma_{1}+\Gamma_{2}}\cdot
(n_{1}^{T}(\varepsilon_1+U_{12})-\nonumber\\&-&n_{1}^{T}(\varepsilon_1+2\cdot
U_{12})\cdot(1+2\cdot A_1))\
\end{eqnarray}

where $\Gamma_i=\Gamma_{ki}+\Gamma_{pi}$. If we introduce tunneling
filling numbers in the absence of Coulomb interaction
$n_{i}^{T}(\varepsilon_i)$, $n_{i}^{T}(\varepsilon_i+U_{ij})$:

\begin{eqnarray}
n_{i}^{T}(\varepsilon)&=&\frac{\Gamma_{ki}f_{k}(\varepsilon)+\Gamma_{pi}f_{p}(\varepsilon)}{\Gamma_{ki}+\Gamma_{pi}}\nonumber\\
\end{eqnarray}

then coefficients $\Phi_{i}$ and $A_{i}$ can be found as:

\begin{eqnarray}
\Phi_{i}&=&\frac{n_{i}^{T}(\varepsilon_i+U_{ii})-n_{i}^{T}(\varepsilon_i+U_{ii}+\cdot
U_{ij})}{3+n_{i}^{T}(\varepsilon_i+2\cdot
U_{ij})-n_{i}^{T}(\varepsilon_i+U_{ii}+2\cdot
U_{ij})}+\nonumber\\&+&\frac{n_{i}^{T}(\varepsilon_i+U_{ii}+2\cdot
U_{ij})}{3+n_{i}^{T}(\varepsilon_i+2\cdot
U_{ij})-n_{i}^{T}(\varepsilon_i+U_{ii}+2\cdot
U_{ij})}\nonumber\\
A_{i}&=&\frac{\frac{1}{2}\cdot
n_{i}^{T}(\varepsilon_i+U_{ij})-\frac{1}{2}\cdot
n_{i}^{T}(\varepsilon_i+U_{ii}+U_{ij})}{3+n_{i}^{T}(\varepsilon_i+2\cdot
U_{ij})-n_{i}^{T}(\varepsilon_i+U_{ii}+2\cdot U_{ij})}-\nonumber\\
&-&\frac{\frac{1}{2}\cdot n_{i}^{T}(\varepsilon_{i}+2\cdot
U_{ij})+\frac{1}{2}\cdot n_{i}^{T}(\varepsilon_{i}+U_{ii}+2\cdot
U_{ij})}{3+n_{i}^{T}(\varepsilon_i+2\cdot
U_{ij})-n_{i}^{T}(\varepsilon_i+U_{ii}+2\cdot U_{ij})}\nonumber\\
\end{eqnarray}

\begin{equation}
F=\left(\begin{array}{c} n_{1}^{T}(\varepsilon_1+U_{11})\cdot n_{1\sigma}\\ n_{2}^{T}(\varepsilon_2+U_{22})\cdot n_{2\sigma}\\ \frac{\Gamma_{1}}{\Gamma_{1}+\Gamma_{2}}n_{1}^{T}(\varepsilon_1+U_{12})n_{2\sigma}+\frac{\Gamma_{2}}{\Gamma_{1}+\Gamma_{2}}n_{2}^{T}(\varepsilon_2+U_{12})n_{1\sigma}\\
\end{array}\right)
\label{matrix_1}
\end{equation}

Pair correlators $K_{ij}$ can be expressed through $n_{i(j)}$ from
equations (\ref{matrix}-\ref{matrix_1}). Substituting the solution
for higher order correlators, obtained from equation (\ref{double})
and (\ref{triple}) to equation (\ref{current_1}) one can find
$<n_{i\sigma}>$ and finally the tunneling current.

The  determinant of the system (\ref{matrix}) can turn to zero or
even becomes negative for some choice of the parameters and
consequently electron filling numbers of the two-level system can
get negative values at some ranges of applied bias voltage. Such
invalid system behaviour is the result of our approximation because
we neglected the interaction between the two localized electronic
states due to the electron transitions to the continuous spectrum
states in the leads and back. To improve the results it is necessary
to take into account corrections which can be found using next order
perturbation theory in the parameter
$\frac{\Gamma_{i}}{\varepsilon_{i}}$, retaining the terms
$t_{k1}t_{k2}\nu_{0k}c_{2\sigma}^{+}c_{1\sigma}$ in equation
(\ref{1}). In this case the final equations for $n_{i\sigma}$ have
additional nonlinear terms and can be schematically written as:

\begin{eqnarray}
n_{1\sigma}\cdot(A_{11}+\mu_{1}n_{2\sigma}^{2})
+n_{2\sigma}\cdot(A_{12}+\mu_{2}n_{1\sigma}^{2})=n^{T}(\varepsilon_1)\nonumber\\
n_{2\sigma}\cdot(A_{22}+\nu_{2}n_{1\sigma}^{2})
+n_{1\sigma}\cdot(A_{21}+\nu_{1}n_{2\sigma}^{2})=n^{T}(\varepsilon_2)\
\end{eqnarray}

Coefficients $A_{ij}$, $\mu_i$ and $\nu_i$ have rather simple but
cumbersome form and depend only on the tunneling filling numbers and
parameters of the tunneling contact.

In this paper we shall not regarded this case.

\section{Main results and discussion}
\begin{figure*} [h]
\includegraphics[width=160mm]{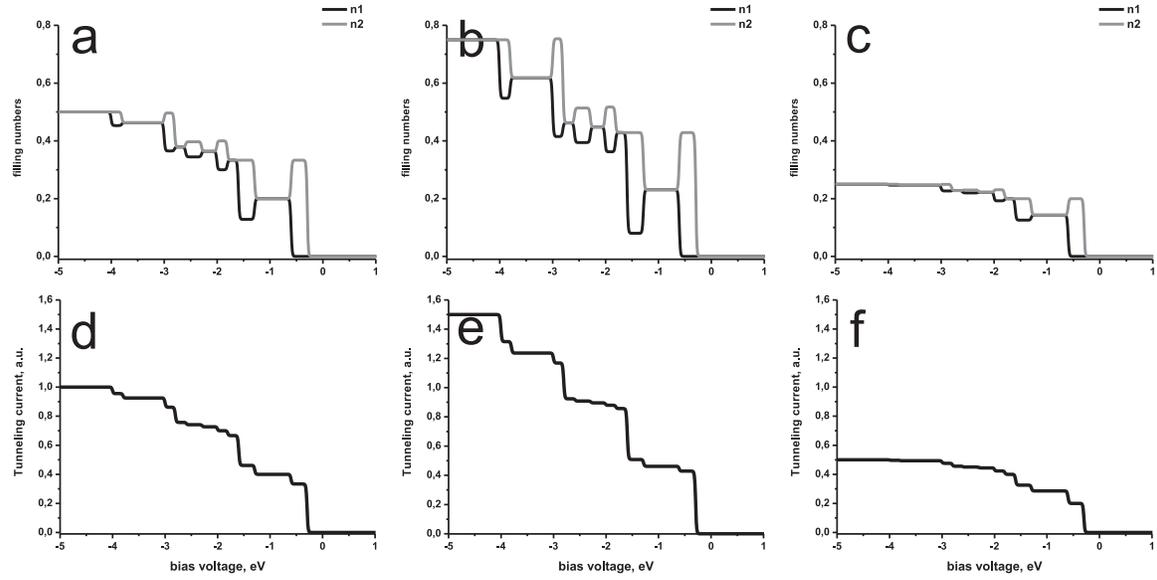}
\caption{Two-level system filling numbers a).-c). and tunneling
current d).-f). as a function of applied bias voltage in the case
when both energy levels are situated above the sample Fermi level.
Parameters $\epsilon_{1}=0.6$, $\epsilon_{2}=0.3$, $U_{12}=1.0$,
$U_{11}=1.4$, $U_{22}=1.5$ are the same for all the figures.
a),d).$\Gamma_{k1}=\Gamma_{k2}=0.01$,
$\Gamma_{p1}=\Gamma_{p2}=0.01$;
b),e).$\Gamma_{k1}=\Gamma_{k2}=0.03$,
$\Gamma_{p1}=\Gamma_{p2}=0.01$;
c),f).$\Gamma_{k1}=\Gamma_{k2}=0.01$,
$\Gamma_{p1}=\Gamma_{p2}=0.03$;} \label{Fig.2}
\end{figure*}

\begin{figure*} [h]
\includegraphics[width=160mm]{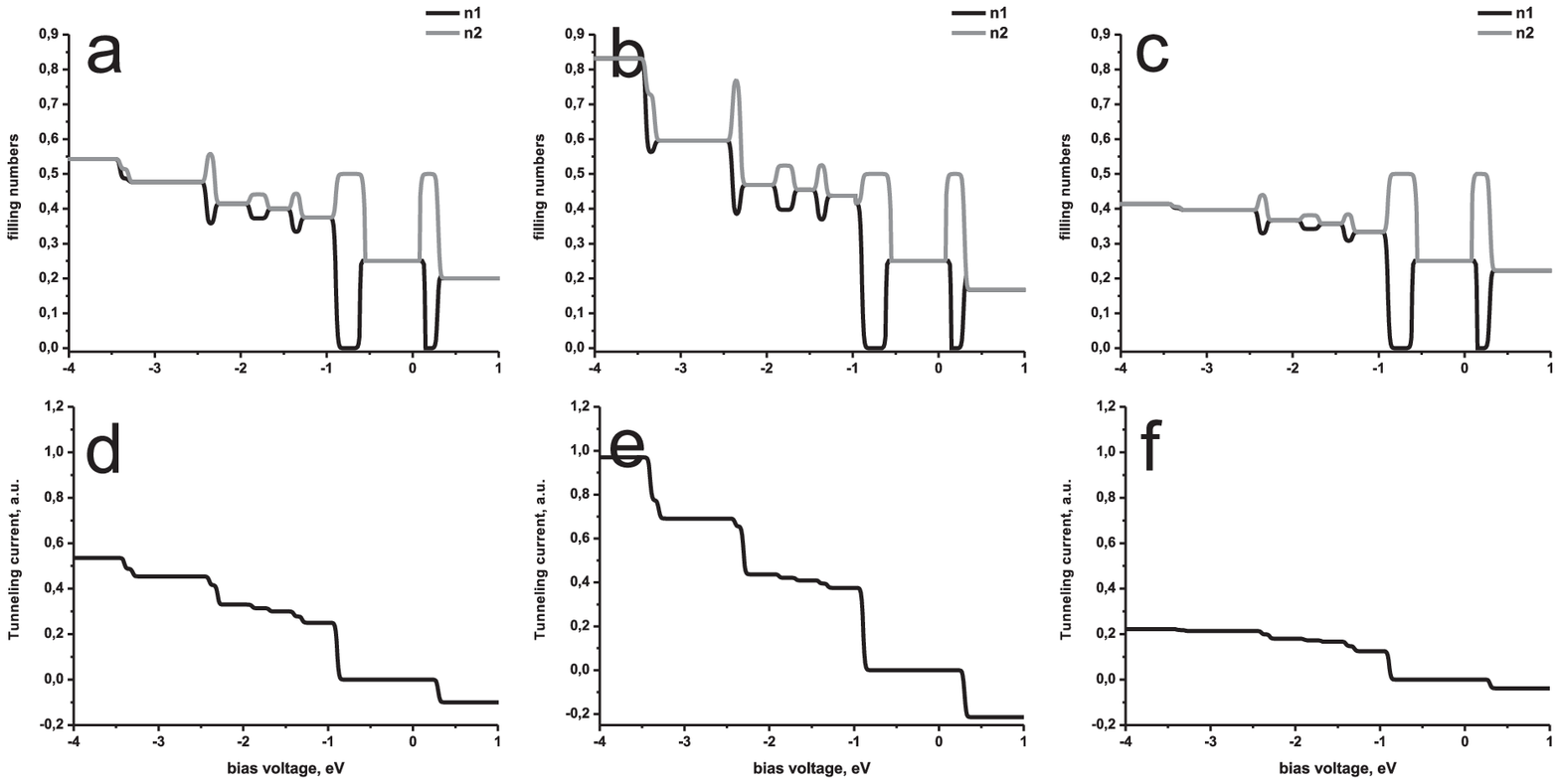}
\caption{Two-level system filling numbers a).-c). and tunneling
current d).-f). as a function of applied bias voltage in the case
when both energy levels are situated below the sample Fermi level.
Parameters $\epsilon_{1}=-0.1$, $\epsilon_{2}=-0.3$, $U_{12}=1.0$,
$U_{11}=1.5$, $U_{22}=1.6$ are the same for all the figures.
a),d).$\Gamma_{k1}=\Gamma_{k2}=0.01$,
$\Gamma_{p1}=\Gamma_{p2}=0.01$;
b),e).$\Gamma_{k1}=\Gamma_{k2}=0.03$,
$\Gamma_{p1}=\Gamma_{p2}=0.01$;
c),f).$\Gamma_{k1}=\Gamma_{k2}=0.01$,
$\Gamma_{p1}=\Gamma_{p2}=0.03$;} \label{Fig.3}
\end{figure*}

\begin{figure*} [h]
\includegraphics[width=160mm]{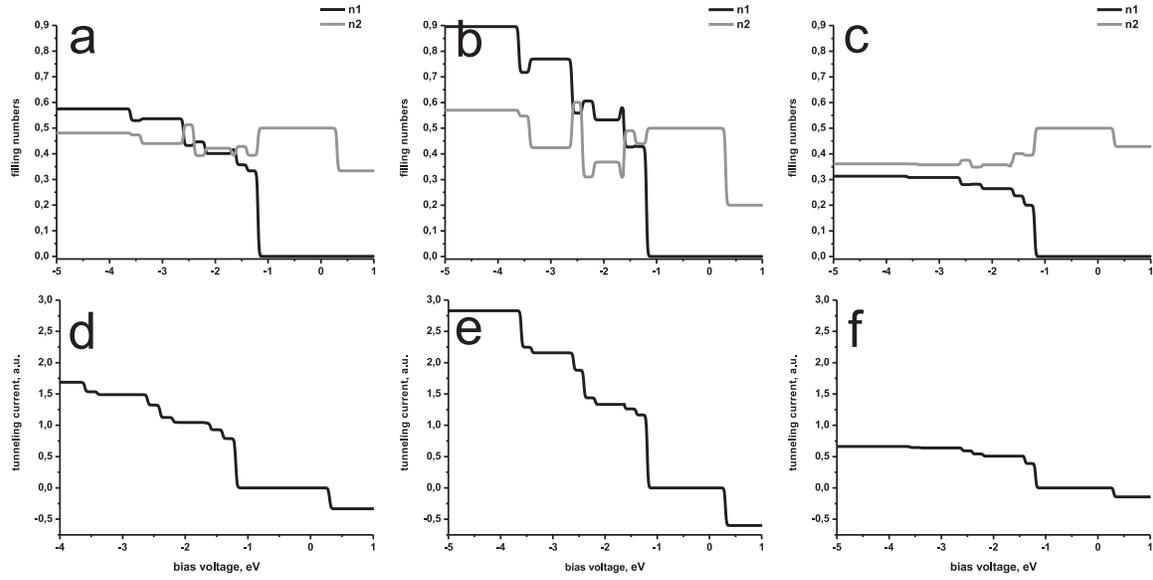}
\caption{Two-level system filling numbers a).-c). and tunneling
current d).-f). as a function of applied bias voltage in the case
when one energy level is situated above and another one below the
sample Fermi level. Parameters $\epsilon_{1}=0.2$,
$\epsilon_{2}=-0.3$, $U_{12}=1.0$, $U_{11}=1.4$, $U_{22}=1.7$ are
the same for all the figures. a),d).$\Gamma_{k1}=\Gamma_{k2}=0.01$,
$\Gamma_{p1}=\Gamma_{p2}=0.01$;
b),e).$\Gamma_{k1}=\Gamma_{k2}=0.03$,
$\Gamma_{p1}=\Gamma_{p2}=0.01$;
c),f).$\Gamma_{k1}=\Gamma_{k2}=0.01$,
$\Gamma_{p1}=\Gamma_{p2}=0.03$;} \label{Fig.4}
\end{figure*}

\begin{figure*} [h]
\includegraphics[width=160mm]{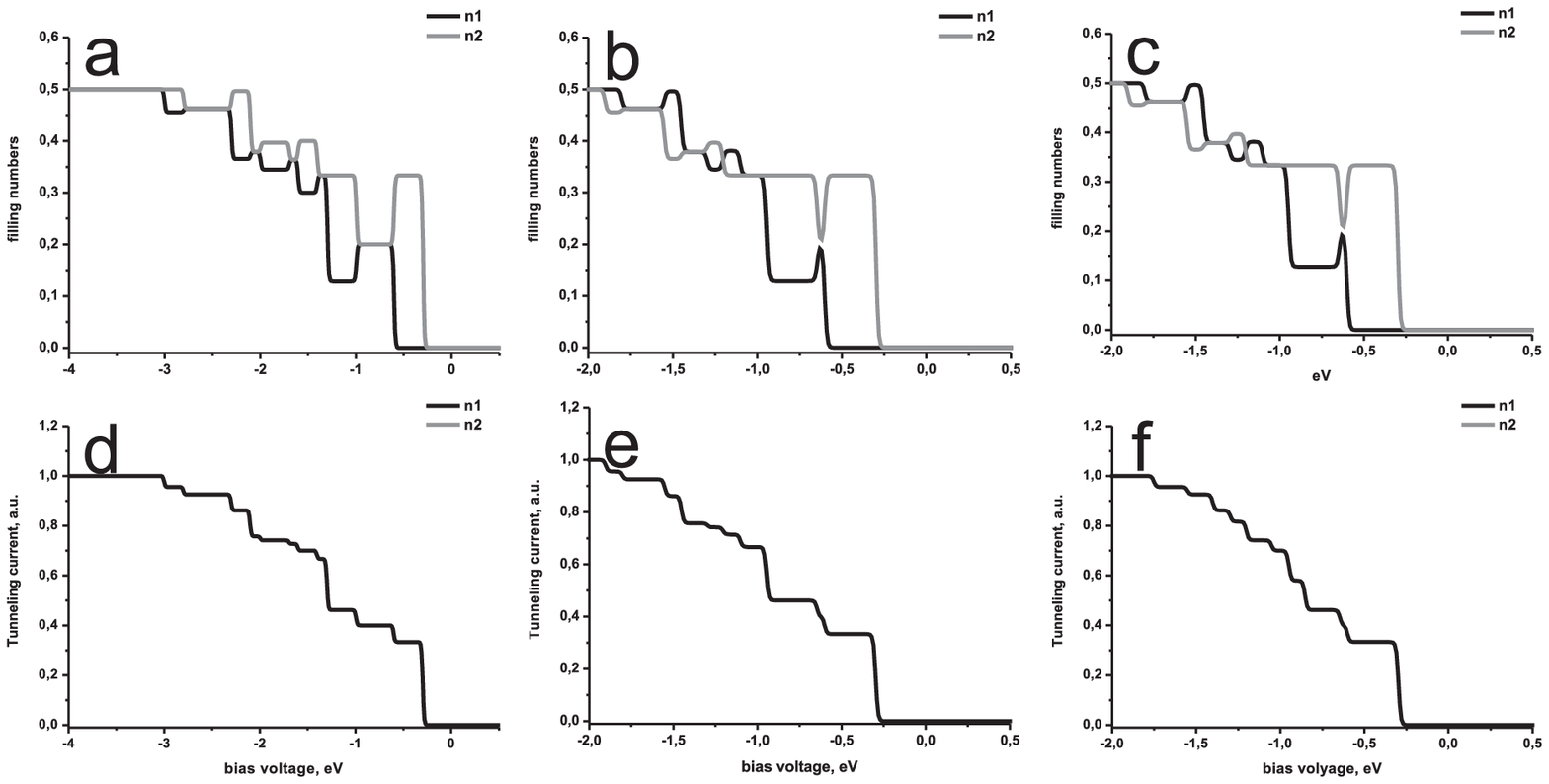}
\caption{Two-level system filling numbers a).-c). and tunneling
current d).-f). as a function of applied bias voltage in the case
when both energy levels are situated above the sample Fermi level.
Parameters $\epsilon_{1}=0.6$, $\epsilon_{2}=0.3$,
$\Gamma_{k1}=\Gamma_{k2}=0.01$, $\Gamma_{p1}=\Gamma_{p2}=0.01$ are
the same for all the figures. a),d).$U_{12}=0.7$, $U_{11}=1.0$,
$U_{22}=1.1$; b),e).$U_{12}=0.35$, $U_{11}=0.5$, $U_{22}=0.9$;
c),f).$U_{12}=0.35$, $U_{11}=0.45$, $U_{22}=0.55$} \label{Fig.5}
\end{figure*}

\begin{figure*} [h]
\includegraphics[width=160mm]{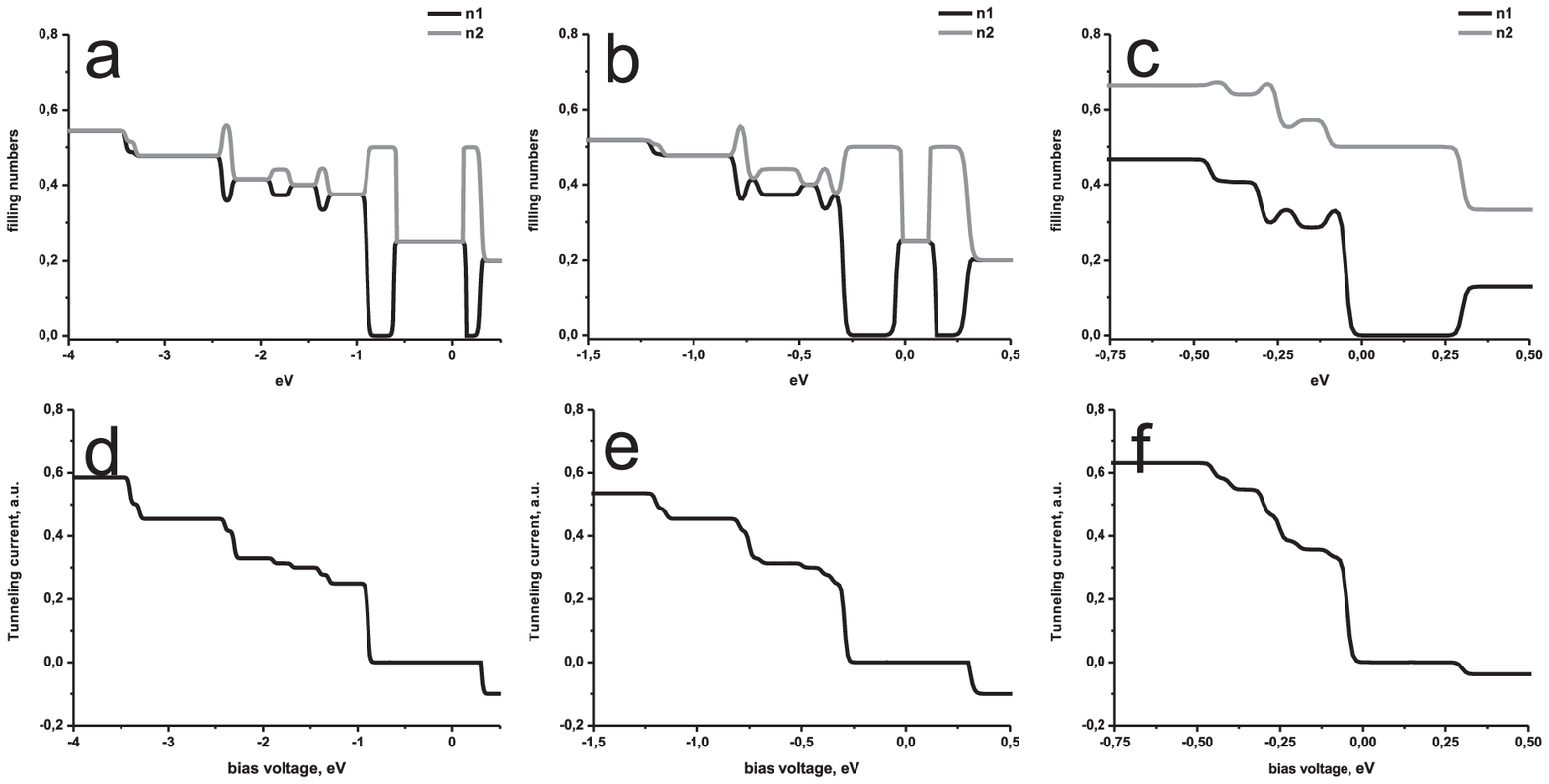}
\caption{Two-level system filling numbers a).-c). and tunneling
current d).-f). as a function of applied bias voltage in the case
when both energy levels are situated below the sample Fermi level.
Parameters $\epsilon_{1}=-0.1$, $\epsilon_{2}=-0.3$,
$\Gamma_{k1}=\Gamma_{k2}=0.01$, $\Gamma_{p1}=\Gamma_{p2}=0.01$ are
the same for all the figures. a),d).$U_{12}=1.0$, $U_{11}=1.5$,
$U_{22}=1.6$; b),e).$U_{12}=0.4$, $U_{11}=0.5$, $U_{22}=0.65$;
c),f).$U_{12}=0.15$, $U_{11}=0.25$, $U_{22}=0.4$}\label{Fig.6}
\end{figure*}

\begin{figure*} [h]
\includegraphics[width=160mm]{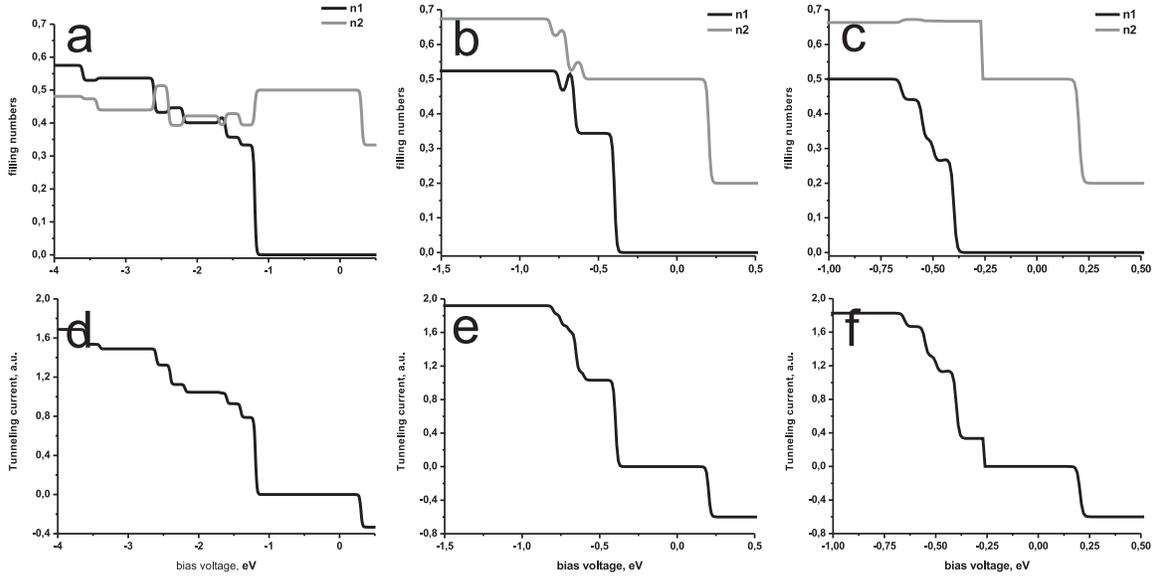}
\caption{Two-level system filling numbers a).-c). and tunneling
current d).-f). as a function of applied bias voltage in the case
when one energy level is situated above and another one below the
sample Fermi level. Parameters $\epsilon_{1}=0.2$,
$\epsilon_{2}=-0.3$, $\Gamma_{k1}=\Gamma_{k2}=0.01$,
$\Gamma_{p1}=\Gamma_{p2}=0.01$ are the same for all the figures.
a),d).$U_{12}=1.0$, $U_{11}=1.4$, $U_{22}=1.7$; b),e).$U_{12}=0.1$,
$U_{11}=0.25$, $U_{22}=0.8$; c),f).$U_{12}=0.1$, $U_{11}=0.15$,
$U_{22}=0.25$}\label{Fig.7}
\end{figure*}

The behaviour of $n_{i\sigma}$ and I-V characteristics strongly
depends on the parameters of the tunneling system. At first let us
analyze the situation in which tunneling rates from both localized
states to the leads are approximately equal $t_{k(p)1}=t_{k(p)2}$.
Figures \ref{Fig.2}-\ref{Fig.7} demonstrate behaviour of filling
numbers and tunneling current obtained from kinetic equations for
the different values of Coulomb energies $U_{ij}$ and various
electron levels location relative to the sample Fermi level in
symmetric $\Gamma_{ki}\sim\Gamma_{pi}$ and asymmetric
$\Gamma_{ki}\ll\Gamma_{pi}$($\Gamma_{ki}\gg\Gamma_{pi}$) tunneling
contact taking into account all order correlators. The bias voltage
in our calculations is applied to the sample. So if both levels are
above(below) the Fermi level all the specific features of charge
distribution and tunneling current characteristics can be observed
at the negative(positive) values of $eV$.

 In the case
of both energy levels situated above (Fig.\ref{Fig.2},
Fig.\ref{Fig.5}) or below (Fig.\ref{Fig.3}, Fig.\ref{Fig.6}) the
sample Fermi level we observe charge redistribution between electron
levels of reentrant character. When applied bias increases two
possibilities for charge accumulation for large values of Coulomb
energies $U_{ij}$ are realized in turn. Charge can be localized on
both electron levels equally $n_{1}=n_{2}$ or mostly accumulated on
the lower energy level ($n_{1}<n_{2}$). In Fig.\ref{Fig.3} there are
two ranges of applied bias where upper level become empty $n_{1}=0$
($\varepsilon_2<eV<\varepsilon_1$ and
$\varepsilon_2+U_{12}<eV<\varepsilon_1+U_{12}$) for large values of
the Coulomb energies. Decreasing of the Coulomb energies leads to
the situation when charge is mostly accumulated on the lower energy
level (Fig.\ref{Fig.6}c), but $n_{1}\neq 0$. In the particular range
of applied bias $\varepsilon_2<eV<\varepsilon_1+U_{12}$ one can find
that the charge is completely localized on the lower energy level:
$n_{1}=0$.

Taking into account all order correlators gives us an opportunity to
investigate tunneling through the two-level system in the case of
small Coulomb energies $U_{ij}\sim\varepsilon_{i(j)}$. Figure
\ref{Fig.5} demonstrates how filling numbers and tunneling current
dependencies change due to decrease of Coulomb energies for the
symmetric tunneling contact $\Gamma_{ki}=\Gamma_{pi}$ (asymmetric
contacts show the same tendencies). We demonstrate the case of both
electron levels localized above the sample Fermi level.

If Coulomb interaction is of the order of single electron energies,
three ranges of applied bias appear, where inverse occupation takes
place: $n_{1}>n_{2}$ (Fig.\ref{Fig.5}b)
($\varepsilon_{2}+2U_{12}<eV<\varepsilon_{1}+U_{11}$,
$\varepsilon_{1}+2U_{12}<eV<\varepsilon_1+U_{11}+U_{12}$ and
$\varepsilon_{1}+U_{11}+2U_{12}<eV<\varepsilon_2+U_{22}+2U_{12}$).
Such situation exists due to the condition that system configuration
with two electrons on the upper level and one electron on the lower
level has lower energy than configuration with one electron on the
upper level and two electrons on the lower level for the parameters
shown in Fig.\ref{Fig.5}b. Further decreasing of the Coulomb
energies (Fig.\ref{Fig.5}c) reduces the effect of inverse occupation
and finally local charge mostly accumulates on the lower energy
level as it should be.

We obtain that the effects of reentrant charge redistribution is
more pronounced for asymmetric contact if tunneling rates to the
sample are lager than tunneling rates to the tip.

It is necessary to mention that without Coulomb interaction one can
find filling numbers for both electron levels to be simple step
functions which correspond to the tunneling filling numbers
$n^{T}(\varepsilon_i)$ shifted from each other on the value
$\varepsilon_1-\varepsilon_2$.

The effect of inverse occupation due to the Coulomb correlations is
more pronounced in a system with electron levels positioned on the
opposite sides of the sample Fermi level.
(Fig.\ref{Fig.4},Fig.\ref{Fig.7}). Without Coulomb interaction, when
$\Gamma_{k(p)1}=\Gamma_{k(p)2}$, the difference of the two levels
occupation numbers
($n_{1}-n_{2}\sim\Gamma_{k1}\Gamma_{p2}-\Gamma_{p1}\Gamma_{k2}$)
turns to zero. Taking into account Coulomb correlations of localized
electrons in the two-level system results in inverse occupation of
the two levels at the wide range of applied bias voltage
(Fig.\ref{Fig.4},Fig.\ref{Fig.7}).

In Fig.(\ref{Fig.4}a,b) there are three ranges of applied bias where
inverse occupation takes place
($\varepsilon_1+U_{11}<eV<\varepsilon_{2}+2U_{12}$,
$\varepsilon_1+2U_{12}<eV<\varepsilon_2+U_{22}+U_{12}$ and
$\varepsilon_{1}+U_{11}+U_{12}<eV$). It is clearly evident
(Fig.\ref{Fig.3}a,b) that when applied bias doesn't exceed the value
$\varepsilon_1+U_{12}$ the whole charge is localized on the lower
energy level ($n_{1}=0$). With the increasing of applied bias
inverse occupation takes place and localized charge redistributes.
The effect of inverse occupation strongly depends on relation
between tunneling rates. It is most pronounced in asymmetric contact
with more strong tunneling coupling to the lead $k$ (sample). But we
have not found inverse occupation if the two-level system strongly
coupled with tunneling contact lead $p$ (tip) (Fig.\ref{Fig.4}c). In
this case with the increasing of applied bias upper electron level
charge increases but local charge still mostly accumulated on the
lower electron level.

Decreasing of the Coulomb energies results in disappearing of the
inverse occupation (Fig.\ref{Fig.7}b,c) and local charge mostly
accumulates on the lower energy level. This clearly demonstrates the
role of Coulomb interaction in described charge distribution
effects.

Tunneling current as a function of applied bias voltage for
different level's positions is depicted in
(Fig.\ref{Fig.2}-Fig.\ref{Fig.7}d-f) (tunneling current amplitudes
are normalized on $2\Gamma_k$). For all the values of the system
parameters tunneling current dependence on applied bias has a step
structure. Height and length of the steps depend on the parameters
of the tunneling contact (tunneling transfer rates and values of
Coulomb energies). If both levels are situated below the Fermi level
(Fig.\ref{Fig.3},\ref{Fig.6}d-f) upper electron level doesn't appear
as a step in current-voltage characteristics but charge
redistribution takes place due to Coulomb correlations.

For approximately equal tunneling rates for both localized levels
current-voltage characteristics are mostly monotonous functions. But
some new peculiarities appear if tunneling rates are essentially
different. In Fig.\ref{Fig.8},\ref{Fig.9} we show some results for
the case $t_{k(p)1}\neq t_{k(p)2}$ . In this case an interplay
between "single electron" nonequilibrium occupation effects and
Coulomb correlation effects exists and at certain bias charge
redistribution is accompanied by negative differential conductivity.

The case of both energy levels situated above the sample Fermi level
is shown in Fig.(\ref{Fig.8}). If the tunneling transfer rate from
the sample to the lower energy level is the largest in the system
and the tunneling transfer amplitude from the lower energy level to
the tip is the lowest one (Fig.\ref{Fig.8}a,c), we see, that local
charge in the system is mostly accumulated on the lower energy
level. Vice versa if the tunneling transfer rate from the sample to
the upper energy level is the largest one and from the upper energy
level to the tip is the lowest in the system (Fig.\ref{Fig.8}b,d),
one can find that local charge is mainly accumulated on the upper
energy level and consequently inverse occupation takes place. But
due to the Coulomb interaction three ranges of applied bias exist
where local charge is mostly localized on the lower energy level
$\varepsilon_2<eV<\varepsilon_1$,
$\varepsilon_2+U_{12}<eV<\varepsilon_1+U_{12}$ and
$\varepsilon_2+U_{22}+U_{12}<eV<\varepsilon_1+U_{11}+U_{12}$.

Inverse occupation also takes place when energy levels are
positioned on the opposite sites of the sample Fermi level
(Fig.\ref{Fig.9}a) or when both energy levels are situated below the
Fermi level (Fig.\ref{Fig.9}b). In any case Coulomb interaction
modifies single electron occupation behavior, changing with applied
bias normal occupation to inverse one or vice versa.

If we look at Fig.(\ref{Fig.9}a), we find several ranges of applied
bias where the charge is distributed differently. These intervals
depend on Coulomb interaction values: the whole charge is
accumulated on the lower energy level ($n_{1}=0$) for
$eV<\varepsilon_1+U_{12}$; inverse occupation exists (local charge
is mostly localized on the upper energy level) for
$\varepsilon_1+U_{12}<eV<\varepsilon_2+U_{22}+U_{12}$ and
$\varepsilon_1+U_{11}+U_{12}<eV$; charge is equally accumulated on
both electron levels $n_1=n_2$ if
$\varepsilon_2+U_{22}+U_{12}<eV<\varepsilon_1+U_{11}+U_{12}$.

If both energy levels are situated below the Fermi level
(Fig.\ref{Fig.9}b) there are similar ranges of applied bias in which
charge is distributed differently (equally for
$\varepsilon_1<eV<\varepsilon_2+U_{12}$, inverse occupation if
$\varepsilon_1+U_{12}<eV<\varepsilon_2+U_{22}+U_{12}$ and
$\varepsilon_1+U_{11}+U_{12}<eV$ and so on).

The appearance of negative conductivity regions is the most
essential feature of the tunneling characteristics, depicted in
Figs.(\ref{Fig.8}c,d, \ref{Fig.9}c,d ). We want to stress once more
that formation of negative conductivity is an interplay between
non-equilibrium effects connected with the tunneling current and
Coulomb correlations.

\section{Conclusion}
      We investigated tunneling through the two-level system with strong Coulomb interaction between localized electrons taking into
account all order correlators of local electron density. It was
shown that charge redistribution between electron states is strongly
governed by the Coulomb correlations and is of reentrant type.
Electron filling numbers dependence on applied bias becomes quite
different from that for non interacting electrons. Existence of the
charge redistribution effects means that adjusting the applied bias
one can control spatial redistribution of localized charges. So wide
possibilities for local charge accumulation and charge switching
exist for such systems

Besides this, at certain values of Coulomb interaction of localized
electrons one can obtain correlation induced inverse occupation of
the two-level system in different ranges of applied bias. Inverse
occupation is mostly pronounced in asymmetric contacts with
different tunneling rates to the sample and to the lead, and when
one energy level lies below the Fermi level and another one - above.

Changing the parameters of the tunneling contact (tunneling rates of
each level to the leads) we can observe interplay between two
mechanisms responsible for non-equilibrium occupation of each level:
tunneling current induced inverse occupation of two-level system at
particular ratio between tunneling rates (which exists in the
absence of Coulomb interaction) and inverse occupation connected
only with Coulomb interaction of localized electrons.

\begin{figure*} [h]
\centering
\includegraphics[width=120mm]{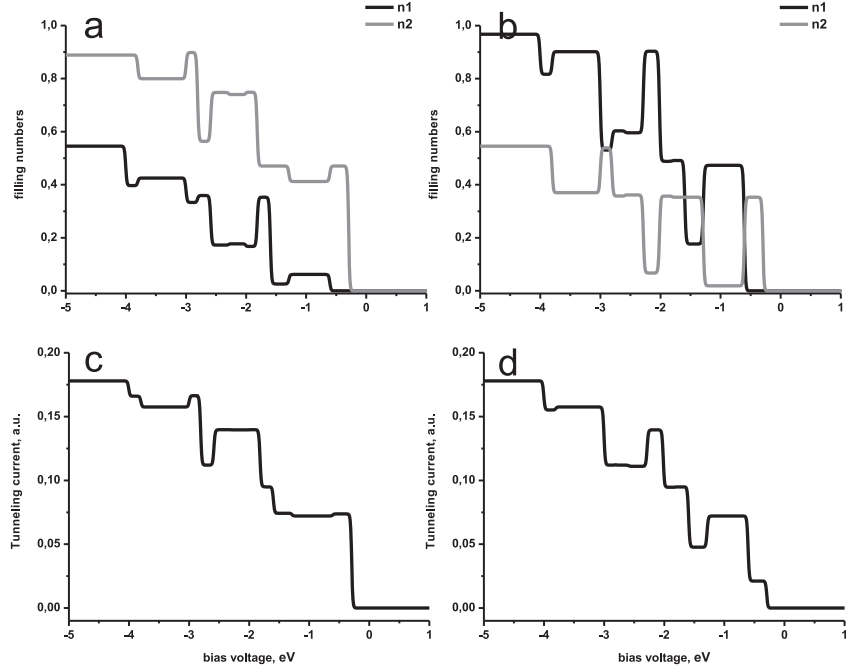}
\caption{Two-level system filling numbers a).-b). and tunneling
current c).-d). as a function of applied bias voltage in the case
when both energy levels are situated above the sample Fermi level
for different values of tunneling rates. Parameters
$\epsilon_{1}=0.6$, $\epsilon_{2}=0.3$, $U_{12}=1.0$, $U_{11}=1.5$,
$U_{22}=1.6$ are the same for all the figures.
a),c).$\Gamma_{k1}=0,06$, $\Gamma_{p1}=0,05$, $\Gamma_{k2}=0,15$,
$\Gamma_{p2}=0,005$; b),d).$\Gamma_{k1}=0,15$, $\Gamma_{p1}=0,005$,
$\Gamma_{k2}=0,06$, $\Gamma_{p2}=0,05$}\label{Fig.8}
\end{figure*}

\begin{figure*} [h]
\centering
\includegraphics[width=120mm]{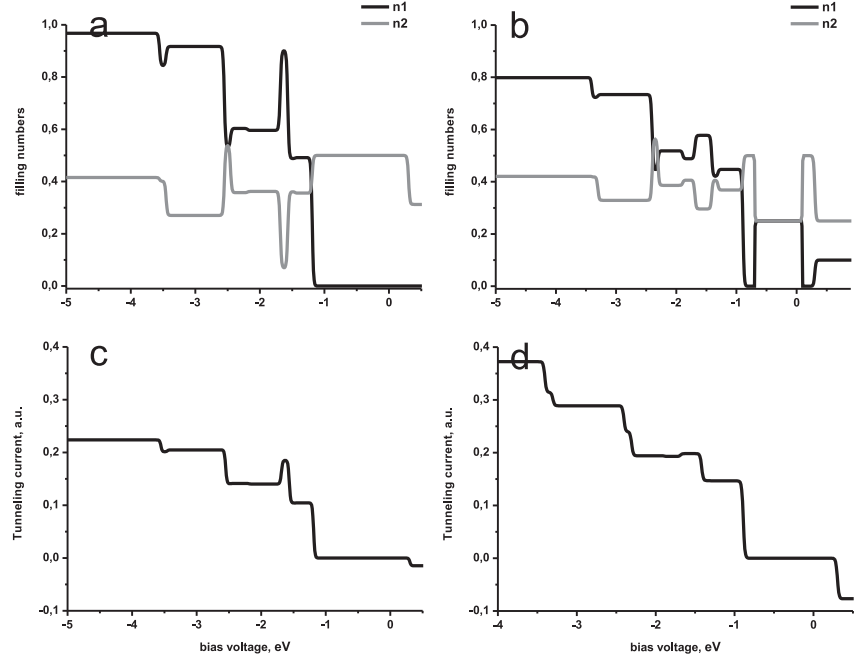}
\caption{Two-level system filling numbers a).-b). and tunneling
current c).-d). as a function of applied bias voltage in the case
when one energy level is situated above and another one below a),c).
and both energy levels are situated below b),d). the sample Fermi
level for different values of tunneling rates. Parameters
$\Gamma_{k1}=0,15$, $\Gamma_{p1}=0,005$, $\Gamma_{k2}=0,06$,
$\Gamma_{p2}=0,05$ are the same for all the figures. a),c).
$\epsilon_{1}=0.2$, $\epsilon_{2}=-0.3$, $U_{12}=1.0$, $U_{11}=1.4$,
$U_{22}=1.7$; b),d).$\epsilon_{1}=-0.1$, $\epsilon_{2}=-0.3$,
$U_{12}=1.0$, $U_{11}=1.5$, $U_{22}=1.6$ }\label{Fig.9}
\end{figure*}

We revealed that for some parameter range system demonstrates
negative tunneling conductivity in certain ranges of applied bias
voltage. Negative tunneling conductivity is revealed in asymmetric
case $\Gamma_{ki}\neq\Gamma_{pi}$ (Fig.\ref{Fig.8} and
Fig.\ref{Fig.9}) and is more pronounced if both energy levels are
situated above the Fermi level. When energy levels are situated on
the opposite sites of Fermi level negative tunneling conductivity is
much weaker and when both of them are positioned below the Fermi
level it is negligible.

Support from RFBR and RAS Programs is acknowledged.

\pagebreak

\end{document}